\documentclass{aa}
\usepackage{graphics}
\input psfig

\begin{document}

\thesaurus{12.03.1; 12.03.3; 12.04.2, 12.05.1}

\title{Can the reionization epoch be detected 
 as a global signature in the cosmic background?}

\titlerunning{Detection of the Reionization Epoch}

\author{P.A. Shaver\inst{1} 
	\and
	R.A. Windhorst\inst{2} 
	\and
	P. Madau\inst{3} 
	\and
	A.G. de Bruyn\inst{4}}

\authorrunning{P.A. Shaver et al.}

\offprints{P.A. Shaver (pshaver@eso.org)}

\institute{
	European Southern Observatory, Karl-Schwarzschild-Str. 2,
	D-85748 Garching bei M\"{u}nchen, Germany
\and
	Arizona State University, Dept. of Physics \& Astronomy, 
	Tempe, AZ 85287-1504, U.S.A.
\and
	Space Telescope Science Institute, 3700 San Martin Drive,
	Baltimore, MD 21218, U.S.A.
\and
	Netherlands Foundation for Research in Astronomy, Postbus 2,
	NL-7990 AA Dwingeloo, The Netherlands \\
	and Kapteyn Astronomical Institute, 
	Postbus 800, NL-9700 AV Groningen, The Netherlands
}

\date{Received 5 October 1998/ Accepted ~~~~~~~~~~~~~~~~~}

\maketitle

\begin{abstract}
The reionization of the Universe is expected to leave a signal in the 
form of a sharp step in the spectrum of the sky. If reionization occurs at 
$5 \la z_{\rm ion} \la 20$, a feature should appear in the radio sky at $70\la 
\nu \la 240$ MHz due to redshifted \ion{H}{i} 21-cm line emission, 
accompanied by another feature in the optical/near-IR at $0.7\la \lambda \la
2.6\,\mu$m due to hydrogen recombination radiation. The expected amplitude 
is well above fundamental detection limits, and the sharpness of the feature 
may make it distinguishable from variations due to terrestrial, 
galactic and extragalactic foregrounds. 

Because this is essentially a continuum measurement of 
a signal which occurs over the whole sky, relatively small telescopes 
may suffice for detection in the radio. In the optical/near-IR, a 
space telescope is needed with the lowest possible background conditions, 
since the experiment will be severely background-limited.

\keywords{cosmology: cosmic microwave background -- early Universe --
diffuse radiation -- observations}

\end{abstract}

\section{Introduction} 

The epoch of reionization marked the end of the ``dark ages'' during which 
the ever-fading primordial background radiation cooled below 3000 K and shifted
first into the infrared and then into the radio. Darkness persisted until
early structures collapsed and cooled, forming the first stars and quasars 
that lit the universe up again. During this epoch the volume filling factor
of ionized hydrogen (\ion{H}{ii}) increased rapidly. It is a generic feature 
of theoretical models and three-dimensional numerical simulations that 
reionization occurred at $5 \la z_{\rm ion} \la 20$ in most cold dark
matter (CDM) cosmogonies (Gnedin \& Ostriker 1997; Haehnelt \& Steinmetz 
1998; Cen 1998; Haiman \& Loeb 1999).\footnote{Note, however, that while
simulations are able to track the formation and merging of dark matter 
halos and the subsequent baryonic infall, they are much less able to predict the
efficiency and rate of radiation emission from gravitationally collapsed 
objects.}~  
We know reionization took place before $z\approx 5$, as evidenced by 
the lack of hydrogen continuum absorption in the spectra of high-redshift 
quasars (Schneider et al. 1991) and galaxies (Franx et
al. 1997). It is unlikely that reionization occurred at $z\ga 50$, 
for in that case the level of degree scale anisotropy in the CMB would
be lower than observed on ten degree scales (e.g. Knox et al. 1998). 
Tilted and neutrino dominated CDM models give $z_{\rm
ion}\simeq$5--8, and open and $\Lambda$-dominated models give $z_{\rm
ion}\simeq$7--30 (Cen 1998).

Moreover, since the characteristic distance between the sources 
that ionized a rapidly recombining intergalactic medium (IGM) was much 
smaller than the Hubble radius, the transition from 
\ion{H}{i} to \ion{H}{ii} was quite abrupt; the overlapping 
of the \ion{H}{ii} regions surrounding
the first star clusters and miniquasars was completed on a timescale much 
shorter than the Hubble time. The reionization epoch occurred almost
as a ``phase transition'' of the Universe. From the observational
point of view, this poorly constrained epoch is among the most important 
unknown quantities in cosmology. 

Various techniques for probing the history of the transition from a neutral
IGM to one that is almost fully ionized have been proposed in the literature.
One way is to look for the 21-cm hyperfine line of neutral hydrogen, 
redshifted to frequencies in the range 70--240 MHz (Madau et al. 
1997; Gnedin \& Ostriker 1997; see also Swarup 1984; Swarup \& Subrahmanyan 
1987; Scott \& Rees 1990). Prior to the reionization epoch, the neutral gas
that had not yet been engulfed by an \ion{H}{ii} region may be seen 
as 21-cm line 
emission if a mechanism existed that decoupled the spin temperature from the 
CMB temperature. Physical mechanisms  that would produce a 21-cm signature
are Ly$\alpha$ coupling of the spin temperature to the kinetic temperature 
of the gas, preheating by soft X-rays from collapsing dark matter halos, and
preheating by ambient Ly$\alpha$ photons (Madau et al. 1997). 

So far the emphasis has been on detecting fluctuations due to individual large
\ion{H}{i} concentrations. This is very difficult in the presence of a
strong fluctuating foreground emission; the \ion{H}{i} lines from
individual concentrations must be resolved, and this requires large 
telescopes with high sensitivity and spectral and angular resolution. 
Alternatively, if reionization occurred abruptly and the fraction of 
neutral hydrogen underwent a drop of a factor of 10$^3$ in about a tenth 
of a Hubble time as predicted ({\it e.g.} Gnedin \& Ostriker
1997; Baltz et al. 1998), it is conceivable that a global 
signature may be detectable with telescopes of modest size in the 
extragalactic background spectrum at meter wavelengths. This is in
principle a much simpler measurement. Some existing 
ground-based telescopes may be suitable, or, if terrestrial interference 
is too severe, a dedicated space project might be considered. A 
complementary technique is to look for a signal in the optical/near-IR
background due to hydrogen recombination radiation from the
reionization epoch (Baltz et al. 1998).

The prospects for detecting a global signature of the reionization epoch are
considered below. In the following we shall denote the present-day Hubble
constant as $H_0=100\,h$ km s$^{-1}$ Mpc$^{-1}$.

\section{Reionization signature} 

Numerical N-body/hydrodynamic simulations of structure formation in a strongly
clumped IGM 
have started to provide a picture for the origin of intervening 
absorption systems, one of an interconnected network of sheets and filaments, 
with virialized systems located at their points of intersection.  
The gas clumping factor rose above unity when the collapsed fraction of 
baryons became non-negligible, i.e. at $z\la 20$, and grew to a few tens at 
$z\approx 8$ (Gnedin \& Ostriker 1997). This made the volume-averaged gas 
recombination time, 
\begin{eqnarray}
\nonumber \bar{t}_{\rm rec} = [1.17\bar{n}_{\rm H} \alpha_B\,C]^{-1} 
= 0.08\, {\rm Gyr} 
\left({\Omega_b h^2 \over 0.02}\right)^{-1} \\
\left({1+z\over 9}\right)^{-3} 
C_{10}^{-1}, 
\end{eqnarray}
shorter than that for a uniform IGM and shorter than the the Hubble
time at that epoch.
In the above expression, $\bar{n}_{\rm H}$ is the mean hydrogen density of the 
expanding IGM, 
$\bar{n}_{\rm H}(0)=1.7\times 10^{-7}$ $(\Omega_b h^2/0.02)$ cm$^{-3}$, 
$\alpha_B$ is the recombination coefficient to the excited states of hydrogen,
and $C\equiv \langle n_{\ion{H}{ii}}^2\rangle/\bar{n}_{\ion{H}{ii}}^2$ is the 
ionized hydrogen clumping factor.

When an isolated point source of ionizing radiation turns on, the ionized
volume initially grows in size at a rate fixed by the emission of UV photons,
and an ionization front separating the \ion{H}{ii} and \ion{H}{i}
regions propagates
into the neutral gas. The evolution of an expanding \ion{H}{ii} 
region is governed by the equation
\begin{equation}
{dV_I\over dt}-3HV_I={\dot N_{\rm ion}\over \bar{n}_{\rm H}}-{V_I\over
\bar{t}_{\rm rec}}, \label{eq:dVdt}
\end{equation}
(Shapiro \& Giroux 1987), where $V_I$ is the proper volume of the 
ionized zone, $\dot N_{\rm ion}$ is the number of ionizing photons 
emitted by the central source per unit time, and $H$ is the Hubble 
constant. Across the ionization front the degree of
ionization changes sharply on a distance of the order of the mean free
path of an ionizing photon. When $\bar{t}_{\rm rec}\ll t$, the growth 
of the \ion{H}{ii} region is slowed down by recombinations in the 
highly inhomogeneous IGM, and its evolution
can be decoupled from the expansion of the universe. As in the static
case, the ionized bubble will fill its time-varying Str\"omgren sphere
after a few recombination timescales,
\begin{equation}
V_I={\dot N_{\rm ion}\bar{t}_{\rm rec}\over \bar{n}_{\rm H}}
(1-e^{-t/\bar{t}_{\rm rec}}). \label{eq:V}
\end{equation}
While the volume that is actually ionized depends on the luminosity of the
central source, the time it takes to produce an ionization-bounded
region is only a function of $\bar{t}_{\rm rec}$.

At early times, the characteristic distance between the sources
that ionized a rapidly recombining IGM was much
smaller than the Hubble radius. The phase transition from
\ion{H}{i} to \ion{H}{ii} was therefore quite abrupt. When the 
\ion{H}{ii} regions surrounding
the first star clusters and miniquasars overlapped, the fraction of neutral 
hydrogen dropped by a few orders of
magnitude in about a tenth of the Hubble time at that epoch, 
the continuum optical
depth decreased suddenly and the background ionizing intensity underwent a
drastic increase in a very small redshift interval (Gnedin \& Ostriker 1997;
Baltz et al. 1998). The redshift of this event is expected to
be essentially the same in all directions. It is this abrupt change 
in \ion{H}{i} absorption that flags the epoch of reionization and may be 
observable at radio and optical/near-IR frequencies, as shown below.

\section{Detectability of an \ion{H}{i} edge}

The intergalactic medium prior to the epoch of full reionization should be 
detectable in 21-cm line radiation. In the absence of a decoupling mechanism, 
the spin temperature of neutral hydrogen would go to equilibrium with the CMB,
and no emission or absorption relative to the CMB would be detected. The 
collapse and cooling of the first non-linear structures had a twofold effect:
(1) it preheated the IGM to temperatures $\sim 150$ K, above that 
of the CMB; and (2) through scattering by Ly$\alpha$ photons it provided 
a mechanism to couple the spin temperature to the kinetic temperature of 
the gas (Madau et al. 1997). A patchwork -- due to large-scale 
structure and 
non-uniform heating and coupling -- of 21-cm line emission would result,  
a signal which would have disappeared after reionization. A signature 
(``step'') in the continuum spectrum of the radio sky, at a frequency of 
$\sim 70-240$ MHz for $z_{\rm ion}$ in the range 5--20,  will therefore 
flag the reionization epoch.

To illustrate the basic principle of the observations we propose, consider a
neutral IGM with spin temperature $T_S\gg T_{\rm CMB}$. 
In an Einstein-de Sitter universe ($\Omega_0=1$, $\Omega_\Lambda=0$), its 
intergalactic optical depth at $21(1+z)\,$cm along the line of sight, 
\begin{equation}
\tau(z) \approx 10^{-3}h^{-1}\left({T_{\rm CMB}\over T_S}\right)
\left({\Omega_bh^2 \over 0.02}\right) (1+z)^{1/2},
\end{equation}
will typically be much less than unity. The experiment envisaged
consists in its simplest form of
two measurements separated in frequency such that the one at shorter 
wavelengths detects no line feature because all hydrogen has been reionized.
The differential antenna temperature across the ``step'', as observed at 
Earth, will be
\begin{eqnarray}
\nonumber \Delta T=(1+z)^{-1} (T_S-T_{\rm CMB}) (1-e^{-\tau}) ~~~~~~~~~~~~ \\
\approx (0.01~{\rm K})
h^{-1} \left({\Omega_bh^2\over 0.02}\right) \left({{1+z_{ion}}\over
9}\right)^{1/2}, \label{eq:dT}
\end{eqnarray}
where $z_{ion}$ is the redshift of the transition epoch. Note that in a universe 
with $\Omega_0\ll 1$, $\Delta T$ increases more nearly linearly with
$(1+z_{ion})$, and the numerical coefficient in Eq. (5) may be larger by up 
to a factor $3[(1+z_{ion})/9]^{1/2}$. Depending on the cosmology, $\Delta T$ will 
increase toward low frequencies at a rate between $\nu^{-1/2}$ and $\nu^{-1}$. 
Hence, the expected cosmological signal will be an edge of more than
$\sim$ 0.02 K (for $h=0.5$) at 70--240 MHz superimposed on the 
2.73 K cosmic background 
radiation. Fig.\ 1 shows the total cosmological signal, comprised of the 
cosmic background radiation with the reionization step superimposed.

\subsection{Fundamental sensitivity limits}

The fundamental limitation on the sensitivity achievable is set by the
intensity of the galactic and extragalactic foregrounds. Modern
receivers are such that the system temperature should be largely 
dominated by these extraterrestrial signals. The limiting sensitivity
is thus
\begin{equation}
\Delta T_{\rm min} \sim \frac{T_{\rm sys}}{\sqrt{\delta\nu t}},
\end{equation}
where $T_{\rm sys}$ is the total system temperature, $\delta\nu$ is 
the bandwidth, and $t$ is the integration time. For $T_{\rm sys}$ = 150K 
(the coldest regions of the sky at 150 MHz) with a bandwidth of 5 MHz 
and integration time 24 hours, $\Delta T_{min}$ $\sim$ 0.0002 K. 
The reionization step predicted by Eq. (5) would be $\sim$ 100$\sigma$, 
independent of telescope size. In a real measurement the system 
temperature would of course be higher than this fundamental limit of 150K, 
but clearly sensitivity is not an issue, and the challenge concerns 
signal contamination and calibration, as discussed below.

\begin{figure}[h]
\centering
\vspace{-22mm}
\hspace*{-3.2cm}
\psfig{figure=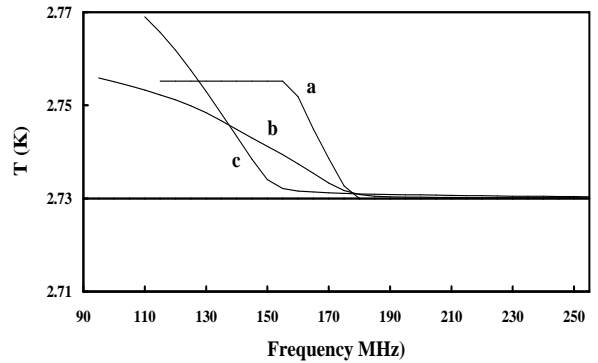,height=12cm,width=8cm,angle=-90.0}
\vspace{-4.5cm}
\caption[]{\small Expected brightness temperature of the cosmic background 
in the vicinity of the \ion{H}{i} reionization edge as a function of observing 
frequency. Three cases are shown for the \ion{H}{i} step, (a): the
case from Eq. (5) with $z_{ion}$ = 7, $\Delta z$ = 1 and 
$h=0.7$, (b) and (c): revised results provided
by N. Gnedin (private communication) from the simulations by 
Gnedin \& Ostriker (1997) and Baltz et al. (1998)
case A respectively, with $\Omega_M=0.35$, $\Omega_\Lambda=0.65$, and $h=0.7$.
}
\end{figure}

\subsection{Contamination by galactic and extragalactic foregrounds} 

The difficulty posed by the galactic and extragalactic foregrounds is
that they can be complex, both in frequency and position. The
question is whether one can in principle extract a 0.02K step from this 
much stronger varying continuum. 

All-sky maps at 150 and 408 MHz are presented in Landecker \& Wielebinski 
(1970) and Haslam et al. (1982). 
The regions of lowest brightness temperatures 
occur in large, relatively smooth regions 
away from the galactic plane and galactic loops. This emission is 
comprised of four components: galactic synchrotron emission 
($\sim$ 70\% at 150 MHz), galactic thermal (free-free) emission ($\sim$ 1\%), 
the integrated emission from extragalactic sources ($\sim$ 27\%), and
the 2.73 K cosmic background itself. \

The galactic nonthermal synchrotron emission is dominant at these low
frequencies, and has been studied since the pioneering days of radio 
astronomy. Its spectrum is close to a featureless power law,
although there are gradual variations in the spectral index with
position on the sky and with frequency. The spectral index is smallest
in the coldest regions of the sky, away from the galactic plane and galactic 
loops. Variations appear to be mainly due to the latter. At 100 MHz 
the spectral index $\beta$ of this component is about --2.55 in cold
sky regions, and it steepens to --2.8 at 1 GHz (Bridle 1967; Sironi 1974; 
Webster 1974; Cane 1979; Lawson et al. 1987; Reich \& Reich 1988; 
Banday \& Wolfendale 1991; Platania et al. 1998). In some regions 
such as the north galactic pole this steepening of the 
spectrum appears to be considerably less pronounced (Bridle 1967).

There may also be a dispersion in the spectral index at each 
position on the sky, due to distinct components along the line of sight. 
Some indication of this dispersion is given by point-to-point variations 
of the spectral index over the sky (Landecker \& Wielebinski 1970; 
Milogradov-Turin 1974; Lawson et al. 1987; Reich \& Reich 1988; 
Banday \& Wolfendale 1990, 1991). It is not large, and appears to be 
smallest at the lower frequencies (Lawson et al. 1987; Banday \& 
Wolfendale 1990). A value of $\sigma(\beta) \sim$ 0.1 is probably 
appropriate at 
100--200 MHz. Further information on small-scale uniformity can be 
provided by simultaneous interferometric data on many short baselines
using telescope arrays such as the WSRT and GMRT. The recently completed 
Westerbork 325 MHz survey of the northern hemisphere (WENSS, de Bruyn et 
al. 1998) already gives some constraints. At high galactic 
latitudes, where the minimum brightness temperature of the diffuse galactic 
emission is about 20 K at 325 MHz, the variations in the brightness
temperature of the galactic foreground emission are typically less than 
about 0.4 K, or about 2\%, on scales ranging from 5--30 arcmin. If
spatial intensity variations are accompanied by spectral variations
(as they would be, if caused by low surface brightness \ion{H}{ii} regions or
supernova remnants), such measurements can be used to set limits on spectral 
variations.

The galactic thermal emission at high galactic latitudes is due to very 
diffuse ionized gas, with total emission measure about 5 pc cm$^{-6}$ for 
$T_{e}$ = 8000K (Reynolds 1990). More recent studies of the 
properties of this gas have been made by Kogut et al. (1996) and
de Oliviera-Costa et al. (1997), who show that it has structure 
correlated with high-latitude dust clouds. This gas is optically thin at 
frequencies above a few MHz, so its spectrum in the region 100--200 MHz is 
well-determined, with a spectral index of --2.1. 

Spectral lines present a related potential contamination. Galactic radio 
recombination lines occur every 1-2 MHz over the frequency range of interest. 
In addition to spontaneous emission from ionized hydrogen and other elements 
in the clouds, there is also stimulated emission (or absorption) against 
the nonthermal background. Peak line intensities can reach 1K or so,
although it turns out that the galactic ridge recombination lines
happen to make the transition from emission to absorption at 100--200
MHz, and so are at a minimum in this range (Payne et
al. 1989). Even a line intensity of 1K becomes diluted to 0.002K in a 
5 MHz band. In any case, spectral lines can be identified and
removed with observations of higher spectral resolution. Observations
such as these may provide the added bonus of discovering or setting limits on
presently unknown spectral lines.

The isotropic emission from extragalactic sources has been estimated both 
directly and from integrated source counts. Bridle (1967) found that this 
component amounts to about 48K at 150 MHz, with a spectral index of 
--2.75. Cane (1979) obtained a value corresponding to about 32K at 150 MHz, 
based on observations at 10 MHz. This is consistent with results from 
integrating the source count data (Willis et al. 1977; Simon 1977; 
Lawson et al. 1987). Simon (1977) estimated that the total 
extragalactic background should turn over between 1 and 2 MHz due to 
synchrotron self-absorption in the individual source components, but
individual sources can have complex frequency structure at much higher
frequencies. In the extreme, spectral indices can range from +2 (sources 
with sharp synchrotron self-absorption turnovers) to --3 (pulsars). Overall, 
the resulting features in the extragalactic background should be small
and slowly varying functions of frequency due to source inhomogeneities and 
the average over many sources. 

This has been simulated using the known spectral properties of 3C sources as 
studied by Kellermann et al. (1969). They classified the spectra of 
almost all 3C sources in the following categories: S (straight -- single 
power law), C-- (convex -- steepening at higher frequencies), C+ (concave 
-- flattening at higher frequencies), and Cpx (complex -- containing 
multiple components). Of the 299 extragalactic sources studied, 42\% are S, 
52\% are C--, 3\% are C+, and 3\% are Cpx. The actual spectral shapes of 37 
3C sources representing all these classes were measured from spectra 
published in Kellermann (1966, 1974), Kellermann et al. (1969), 
Kellermann \& Pauliny-Toth (1969), and Kellermann \& Owen  
(1988). These were then weighted in accordance with the frequency of
occurance of the 
various classes, and co-added to give a representative composite spectrum 
of the extragalactic foreground.

The resulting extragalactic spectrum in the range 50-300 MHz is close to 
a power law of spectral index --2.65 (the index is steeper at higher 
frequencies). The deviation from this power law is a slowly and 
continuously varying function of frequency, with an index ranging from
--2.59 to --2.66 and no sharp features. The smoothness of this spectrum 
based on just 37 sources (see Fig. 3) is reassuring, as vastly more 
sources will contribute to the real 
extragalactic signal, which will be correspondingly more featureless. 
The overall spectrum of the real extragalactic signal will differ from
that computed here, as only the most luminous sources are represented 
in this simulation, but the important conclusion is that the spectrum 
should be smooth and featureless.

\begin{figure}[ht]
\centering
\vspace{-6mm}
\hspace*{-4.5cm}
\psfig{figure=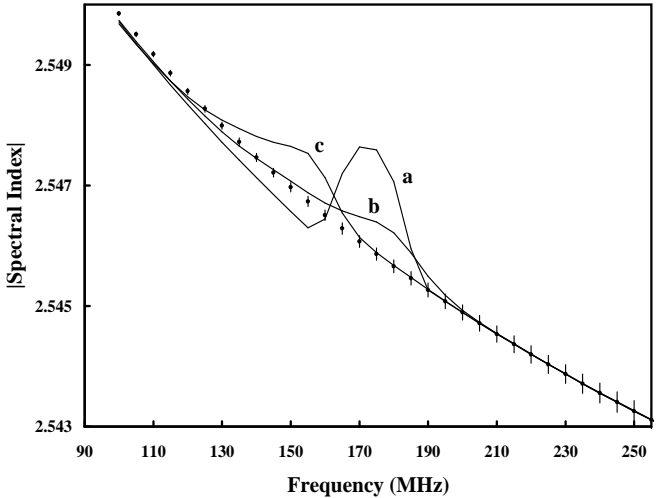,height=12.2cm,width=9cm,angle=-90.0}
\vspace{-20mm}
\vspace{-2.4cm}
\caption[]{\small Spectral index {\it vs.} observing frequency, as 
deduced over 10 MHz intervals from simulated spectra which include the
galactic nonthermal emission with median spectral index $\beta$ = -2.55 
and a Gaussian dispersion around this value with $\sigma$ = 0.1, 
and the galactic thermal emission with $EM$ = 5 pc cm$^{-6}$ and 
$T_{e}$ = 6000K, in addition to the reionization signal (the constant
2.73K cosmic background has been removed). The emission from 
extragalactic sources is omitted in this case. The total 
brightness temperature at 150 MHz is 150K. The points represent the 
case without a reionization step, and the error bars correspond to 
$\pm 5\sigma$ for a 24-hour observation. The lines represent the three
cases shown in Fig.~1.}
\end{figure}

Nevertheless it will still be important to identify and characterize 
the spectral properties of the dominant sources within the main beam. 
Interferometer observations can be made for this purpose. A single source 
with a flux density of 35 mJy at 150 MHz would produce
an antenna temperature increase of 0.01 K for a 45m telescope with 50\% 
main beam efficiency. All sources should therefore be identified down 
to at least this level. The GMRT would be the instrument of choice to conduct 
these high resolution observations. With receivers for 50, 150, 235 
and 325 MHz and a resolution of better than 1' (at 50 MHz), spectra of
all significant sources can be determined in the relevant frequency range.

\begin{figure}[h]
\centering
\vspace{-6mm}
\hspace{-45mm}
\psfig{figure=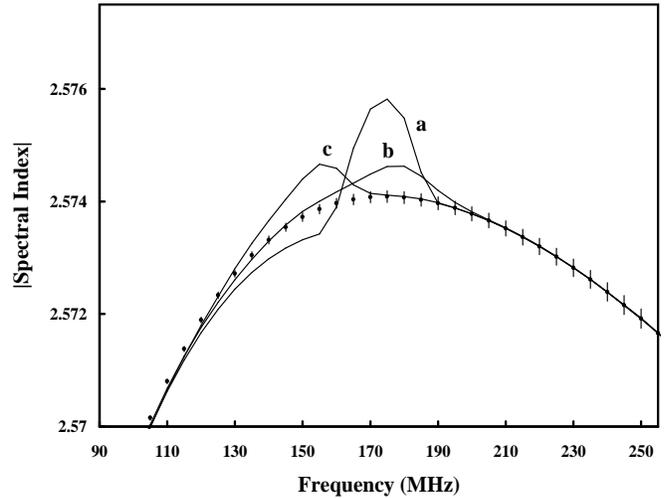,height=12.2cm,width=9cm,angle=-90.0}
\vspace{-4.4cm}
\caption[]{\small Spectral index {\it vs.} observing frequency as 
in Fig. 2, except that in this case the emission from extragalactic
sources is included, represented by the simulation described in the
text based on the actual spectra of 37 3C sources. Again, the 
total brightness temperature at 150 MHz is 150K.}
\end{figure}

Thus, it may be possible to quantify all the foreground contaminants
sufficiently to model them and possibly remove some of them. The 
fact that the reionization signal should
be the same over the whole sky is a great advantage - searches in 
different positions on the sky with different contamination should give 
the same result. A range of different measurements can help greatly in
disentangling these foreground signals. Accurate broadband spectra over 
a wide frequency range can help to determine the spectral behavior of
the galactic nonthermal component, and measurements in adjacent positions 
can constrain the dispersion of the spectral index. High frequency 
measurements can be used to accurately determine the galactic thermal 
component. Interferometer observations can be used to identify regions
devoid of strong extragalactic sources, and to measure the properties 
of the most prominent sources  in the regions chosen. 

The analysis can be done in a number of ways, depending on how effectively 
the different components can be modeled or removed. If the overall spectrum 
could be well determined at frequencies above that of the reionization
step ({\it e.g.} $\nu >$ 237 MHz, corresponding to $z < 5$), a simple 
extrapolation to lower frequencies could be made to find the step. It 
is unlikely, however, that such an extrapolation could be made with 
sufficient reliability. Alternatively, after those components that are 
well-determined have been removed (at least the 2.73 K cosmic 
background and the 
galactic thermal emission), a best fit (power law or low-order polynomial) 
could be made and subtracted or divided from the actual spectrum to reveal 
the step, although the fit may well introduce artifacts that can mask or 
confuse the signal. A natural approach is to make the fit at the two ends 
of the spectrum and interpolate, but this requires a well-behaved spectrum
and assumes that the signal is located near the middle of the
spectrum. Another possibility is to measure the spectral index 
point-by-point along the spectrum, and look for an abrupt jump due to 
the step. 

Figure 2 shows the spectral index measured every 10 MHz from simulated 
spectra containing just the galactic thermal and nonthermal emission 
in addition to the reionization signal. In Fig. 3 the extragalactic 
sources have been included. These two examples give an idea of the range 
of spectral curvature that can be expected. The overall curvature 
is substantial, but the reionization signal can still be seen in all three 
cases because of its relative sharpness.

\begin{figure}[ht]
\centering
\vspace{-8mm}
\hspace{-25mm}
\psfig{figure=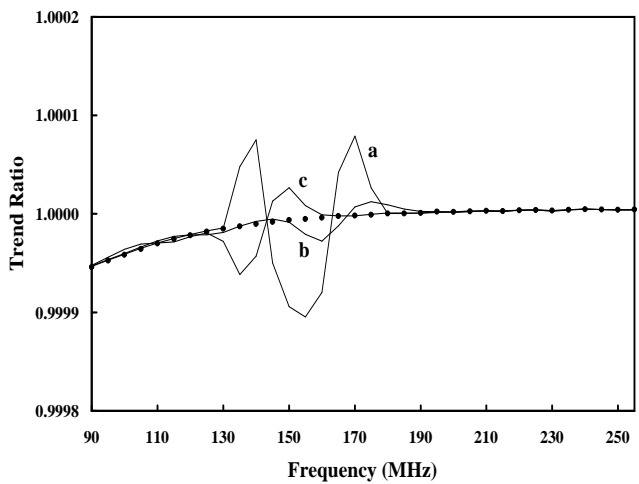,height=11cm,width=7.5cm,angle=-90.0}
\vspace{-35mm}
\caption[]{\small The ``trend ratio'' as described in the text, computed 
for the curves used in Fig. 3, plotted against observing frequency.}
\end{figure}

Another technique that does not involve any fitting procedure (but does 
require high-precision measurements) is trend analysis -- analyzing the 
spectrum for deviations from a smooth trend. Fig. 4 shows the "trend 
ratio", defined here as the ratio of the observed brightness temperature 
at a given frequency to that predicted from an extrapolation of the 
brightness temperatures measured at the three higher frequency points. 
The disruption caused by the reionization signal to the otherwise smooth 
spectrum is obvious.

Clearly, both the magnitude and sharpness of the reionization step are
critical parameters. As mentioned at the beginning of Sect. 3, the 
magnitude of the step will be greater in a low density Universe. 
The sharpness depends on the clumpiness of the Universe at the reionization
epoch and the formation rate of the ionizing sources, and may well be
greater than indicated by curves (b) and (c). Simulations of the 
evolution of structure can give some indication of these parameters, 
but ultimately they have to be determined from observations.

In summary, it appears that the reionization step may be detectable in
principle, even in the presence of contamination due to the galactic 
and extragalactic 
foregrounds. How easily it can be detected, and which method of analysis 
is best, will depend on how well the different contaminating components 
can be measured, modeled and removed. For this purpose, new and
accurate studies of these foreground contaminants will be very important. 
The remaining limitations are then purely technical, and present an 
interesting challenge. Some of the practicalities are discussed below.

\subsection{Observational prospects}

\subsubsection{Telescope requirements} 

As (1) the signal is present over the entire sky, (2) the system
temperature is dominated by the galactic and extragalactic
nonthermal emission, and (3) this is a broadband (rather than line)
observation, this experiment can in principle be done using already 
existing radio telescopes of moderate aperture. The main requirement 
is to achieve highly accurate relative intensity measurements over a 
considerable frequency range (70--240 MHz).

Attractive as the use of a small telescope for this measurement may
be, however, there are significant difficulties. The gain of the
telescope, its beamshape and its (distant) sidelobes are
frequency-dependent, due to matching, feed illumination, feed support
scattering and edge-refraction effects. The beamsize
increases with decreasing frequency, so the regions contributing to the 
galactic and extragalactic emission are different at different
frequencies. This will further complicate the shape of the overall
spectrum, although the effect should vary slowly and smoothly with
frequency, particularly in the directions of lowest galactic emission.
The distant sidelobes may also introduce (varying) signals from regions
of stronger galactic emission. 

The large beam of a small telescope would average over the effects of 
many extragalactic sources, and more sources would be included at
lower frequencies where the beam is larger. Averaging over many
sources could conceivably be an advantage, as long as a small number
of sources did not dominate, but with a larger telescope and smaller
beam it would be easier to observe in regions devoid of strong 
extragalactic sources. Finally, if a strong source such as Cas A is to
be used as a calibrator (see below), a fairly large antenna is
required for the signal from this source to dominate over the
surrounding galactic emission. 

A larger parabolic antenna would solve some of these problems, but 
not all. Some sort of phased array with beam-forming circuitry which 
is still sensitive to emission on the largest scales may have several
advantages. It would be possible to use scaled beams at all frequencies 
and control sidelobes, so that exactly the same area of sky
would be sampled at all frequencies. New arrays of this type, such as
the THousand Element Array (THEA) in The Netherlands and the Square
Kilometer Array (SKA) are presently under consideration (van Ardenne 
1998). 

If terrestrial complications such as radio-frequency interference were 
considered insurmountable, one might consider a small satellite project.
Then there would be only the galactic and extragalactic foreground 
contamination to contend with, as well as calibration (although solar
radiation may also be a significant problem). The antenna size 
would obviously be limited, but even a deployable 10m antenna, giving 
a 10$^{\circ}$ beam at 180 MHz, may be adequate. At these frequencies 
the precision of the antenna surface would not be an issue. While space 
projects are not inexpensive, a space radiotelescope such as this may find 
novel uses, as it would be exploring a new domain in observational 
phase space.

\subsubsection{Calibration options} 

To be specific we will discuss three different calibration strategies,
and mention specific problems for each of them and how they might be solved.
Calibrating the frequency response of the
telescope/receiver combination  is probably {\em the} crucial issue. 
This has to be done to better than a few parts in 10$^{5}$ over a wide 
frequency range. Possibilities (which could be combined) include 
internal loads, or the use of astronomical sources - a very strong 
'featureless' continuum source such as Cas A, or the moon. We discuss
each of these in turn.

\vspace{0.3cm}

\noindent
{\bf a) Internal loads} 

For precise calibration over many frequency channels, internal
broad-band noise sources of the highest quality would undoubtedly 
be the best option. They would have to be extremely stable and very
accurately calibrated as a function of frequency. The properties of
available internal calibration sources will have to be examined
carefully to determine whether this is an option -- until now, 
calibration of this accuracy and stability has not been required at these
frequencies, so this is uncharted territory. Calibration with 
an internal load must also take care of any frequency dependence of 
the load-coupling into the signal path.

\vspace{0.3cm}

\noindent
{\bf b) Using Cas A} 

The alternative is to use astronomical calibration sources. This has
the advantage that the astronomical signal and the calibration signal 
both traverse the same path through the telescope and receiver system 
and therefore share similar gain-frequency effects. Cas A is
the strongest source in the sky at 150 MHz, with a flux density of
13,000 Jy. In order for this source to produce an antenna temperature
well in excess of the local galactic emission (Cas A sits on the
galactic ridge), a telescope of 25m diameter or more would be required. 
The use of such a source may introduce a further complication:
although its overall power-law spectrum is remarkably straight down to
a few MHz (Baars et al. 1977), many small components with a 
range of spectral indices are superimposed, modifying the detailed 
overall spectrum to an extent yet to be determined (Anderson \&
Rudnick 1996, and references therein). Strong extragalactic sources 
such as Cygnus A have more complex, curved  spectra (Baars et al. 
1977), and so may be less suitable as calibrators. 

Probably the best existing telescope to begin with is the GMRT. Cas A 
would produce an antenna temperature of about 4300 K at 150
MHz with the 45m antennas, totally dominating the sky and receiver noise. 
The GMRT has 
receivers over a wide range of frequencies, but not contiguous over the
desired range from 70--240 MHz. Total power measurements, accompanied by
autocorrelation spectra of the input signals to identify data affected
by narrow band interference, are required. Digital spectrometers 
with at least 20-40 MHz bandwidth and several thousand spectral
channels would be needed.

\vspace{0.3cm}

\noindent
{\bf c) Using the moon} 

An advantage of the moon over Cas A is that it moves, and a true 
differencing experiment could therefore be done at exactly the same 
position(s) in the sky, and in the same alt-az position(s), but of 
course not at the same time. All stable contaminating effects
(galaxy, discrete sources, sidelobes etc) will to first and 
perhaps second order be identical and will cancel, leaving only
time-dependent concerns - overall stability and radio-frequency interference.
A true differencing experiment such as this may be very attractive in
coping with the contribution to the system temperature coming in via 
distant sidelobes. It will be hard to limit this contribution to less 
than 50 K at the low frequencies of this experiment. This component may 
have significant frequency structure which could be difficult to model. 
This may be a great advantage of using the moon compared to an 
experiment using Cas A.

The brightness temperature of the moon is about 220 K in this
frequency range (Keihm \& Langseth 1975); it varies
slowly with lunar phase. This is close to the minimum sky temperature at 150
MHz, so if the moon fills the beam it will not increase the antenna
signal by much, if at all. The fact that the moon completely blocks the 
emission behind it (galactic and extragalactic) can be very useful, and 
could help solve the problem of the frequency-dependent beamsize. The moon 
is not a very bright calibration source so it takes a large telescope 
to avoid diluting the signal too much. For the moon to fill 
the beam, we require a telescope of about 200--400 m diameter at the 
frequencies that we are considering. Currently only Arecibo would be 
able to provide a sufficiently narrow beam. A specially designed 
low-frequency phased-array, such as those mentioned in Sect. 3.3.1, may
be required. 

An interesting aspect of a moon calibration experiment is that the
detection of the signal could in principle be done interferometrically.
Because the moon occults the background signal it introduces spatial
structure in the global edge signal. 
Depending on the frequency and the intensity of the galactic 
background, the 220 K moon could be a colder or warmer spot in the sky. 
But the intensity of this negative or positive source contains 
the frequency-dependent global background temperature. 

Of course, the motion of the moon also introduces smearing effects,
which would probably be advantageous. If the 'edge' signal is
universal, as expected, we need not worry that the moon moves by
approximately its 
diameter every hour. Both the sought-after signal and the sidelobes will
be averaged. Since we need to integrate over many (tens of) hours by 
tracking the moon we are effectively sensitive only to a large-scale signal. 
The possibility that emission from the moon, which includes a time-variable 
contribution from scattered solar radiation, may have  weak frequency 
structure will have to be investigated. 

Alternatively
(especially if smaller telescopes are used) even the sun might conceivably 
serve for calibration purposes. Although the solar spectrum has 
considerable frequency structure during bursts, the quiet sun is less 
likely to have permanent spectral features at 70--240 MHz, since 
we are looking well into the ``quiet'' spectrum of the solar chromosphere
superimposed on the Rayleigh-Jeans tail of its 6000 K atmosphere. 
Several issues will have to be further studied before deciding on the 
optimal calibration strategy.

\subsubsection{Terrestrial interference and complications} 

\noindent
Radio-frequency interference (RFI) is of course a major issue for
observations at the frequencies of interest. The World Radio Conference 
had allocated only one protected radio astronomy band in this regime,
at 150--153 MHz, which itself is already severely affected by interference
(Spoelstra 1998). This region of the radio spectrum is subject to local, 
regional and world-wide interference.  Some of these are stationary 
sources of interference such as TV stations, many other in the 
100--200 MHz regime are of a mobile nature such as aeronautical and
military transmitters and beacons. Fortunately, many of these sources 
are strongly time-variable and confined to relatively narrow bands.
The 88--108 MHz band is allocated to FM radio and may be completely 
inaccessible for radio observations of the required accuracy. 
This means that if the \ion{H}{i} 'edge' signal is at redshift greater than 10 
(i.e. at frequencies less than 130 MHz) there may not be a 
reliable spectral baseline left.

It is interesting to consider the role that could be played by
interferometers in this experiment. Such arrays can, with many 
independent monitors of the total power, provide an excellent discriminator 
(filter) for local or external RFI. Because RFI generally does not come 
from the direction of the signal its delay and fringe rate will also be 
different. This then provides a way of detecting and possibly eliminating 
RFI, so using telescopes that are part of an interferometric array may
yield powerful diagnostic capabilities for weak RFI. 

Ionospheric effects are also important at these frequencies. The 
ionosphere not only refracts the signal by a varying amount (easily 1-2 
arcminutes) but there is also refractive focussing and scintillation. 
These scale with frequency in a quadratic way. However, the ionosphere
changes daily, so one can always select the best conditions. The
effects would be of greatest concern for the observations of the discrete 
calibration sources, and should not be important for low-resolution
observations of the frequency dependence of large areas of sky.

\section{Detectability of a Lyman reionization edge}

In this section we discuss the possibility of detecting the reionization edge
via the Lyman lines. We first discuss the nature of this optical/IR signal, 
next its observability in general, and then we conclude by discussing an 
experiment with existing equipment. 

The rate of hydrogen recombinations peaks at $z\simeq z_{\rm ion}$, and this is
the source of the background we seek. Because we want to observe a spectral
signature in the all-sky background, we are only concerned with how energy is
redistributed among different spectral lines. Baltz et al. (1998) showed that
intergalactic absorption brightens some lines while it dims others. At $z\ge
z_{\rm ion}$, the optical depth in the Lyman series is very high. All Lyman
series lines except Ly$\alpha$ are absorbed immediately and redistributed. This
makes Ly$\alpha$ and the Balmer series significantly brighter, because they
receive the energy of the Lyman series. Once reionization occurs, the Lyman
series becomes optically thin and the brightness of Ly$\alpha$ and the Balmer
series drops considerably, producing sharp features in the background
spectrum. 

If reionization occurs abruptly, such a sharp edge may be observable as a
global signal over the entire sky due to Ly$\alpha$ emission at 
z$\simeq$5--20 ($\lambda\simeq$0.7--2.6$\mu$m). This can be seen from the
models of Baltz et al. (1998) shown in Fig. 5. As we look redward, crossing
Ly$\alpha$ around $\sim$1$\mu$m during the reionization event, the brightness
suddenly increases due to the absorption and redistribution of the higher Lyman
series lines. Haiman et al. (1997) calculated a similar effect in which
Ly$\alpha$ is brightened by radiative excitations, but it appears that
recombination gives a significantly stronger signal. Baltz et al. (1998) and
Longair (1995) estimate the strength of the edge to be in the range
$J_{\nu}({\rm Ly}) \simeq 0.3-3\times 10^{-23}$ erg/cm$^2$/s/Hz/sr (these
estimates are probably uncertain by $\ga$~0.5 dex). 

Further recombination features are also present in the spectra in Fig. 5. The
broad humps (most noticeably that blueward of Ly$\alpha$) are due to blends of
lines from the Lyman and Balmer series (Baltz et al. 1998; see also Stancil et
al. 1998 for other possible transitions with lower probability). The positions
of these humps are of course fixed relative to the sharp features. Their shapes
are determined by the evolution of the reionization/recombination history.

\begin{figure}[h]
\centering
\vspace{-2mm}
\resizebox{\hsize}{!}{\includegraphics{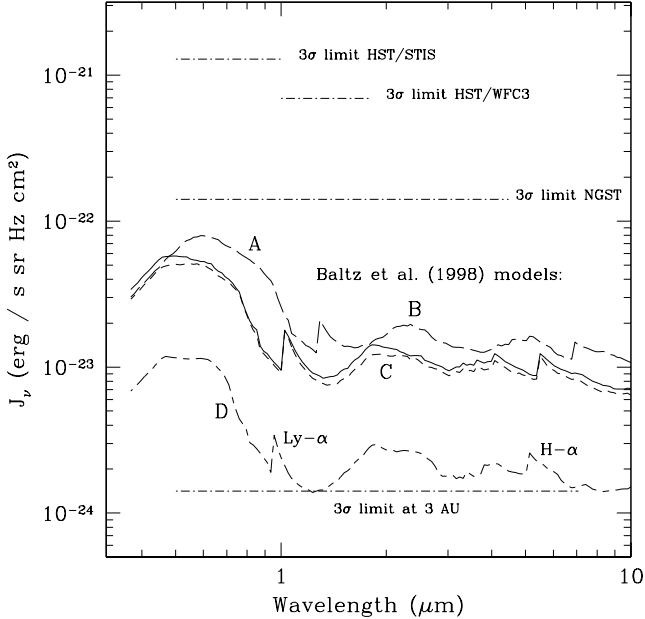}}
\caption[]{\small Predicted spectra of the Lyman reionization edge from Baltz
et al.\ (1998). Models A--D were generated from CDM simulations including
radiative transfer, star-formation, feedback, evolution and chemistry of the
gas, and assume $\Omega_M=0.35$, $\Omega_\Lambda=0.65$, and $h=0.7$. The
sequence A--D has decreasing values of $\Omega_{b}$ (0.055-0.03), simulation
box-size (3--2 h$^{-1}$ Mpc), mass resolution ($10^{7.0}$--$10^{6.0}$
$M_{\odot}$), and number of particles. The amplitude prediction in the models is
probably uncertain by $\ga$~0.5 dex. The most noticeable hydrogen edges,
Ly$\alpha$ and H$\alpha$, are indicated (for model D only). The hump blueward
of Ly$\alpha$ is caused by higher atomic transitions. The dotted curves
indicate the in-principle 3-$\sigma$ limits expected from a day of
integration with HST/STIS, HST/WFC3, NGST, and a long integration on a similar
spacecraft at 3 AU from the sun. The uncertainty in these limits are about
$\sim$0.3 dex.}
\end{figure}

\subsection {Fundamental sensitivity limits}

The sky brightness at these wavelengths is dominated by the zodiacal light. The
Diffuse Galactic Light (DGL), and the Extragalactic Background Light (EBL) also
contribute, but at a lower level. These ``foreground'' emissions determine
the fundamental sensitivity limits for this observation.

The dominant foreground signal is the zodiacal ``background'', which at high
galactic and ecliptic latitudes is well determined by HST in the WFPC2 B, V, I 
bands at $\sim$23.9, 23.00, and 22.45 arcsec$^{-2}$, respectively (Windhorst et
al. 1992, 1994b, 1998), and $\ga$20.9 and $\ga$19.6 mag arcsec$^{-2}$ in the
Near-Infrared Camera/Multi Object Spectrograph (NICMOS) near-IR J and
H bands, respectively (Thompson et al. 1999; Windhorst \& Waddington 1999).
The latter are upper limits to the IR zodiacal sky, since there is some
remaining uncertainty in the global NICMOS dark-current. At 0.55$\mu$m, the
zodiacal sky as measured from low earth orbit corresponds to $J_{\nu} \sim
6.8\times 10^{-19}$erg/cm$^2$/s/Hz/sr, and at 1$\mu$m it corresponds to
3.1$\times 10^{-18}$erg/cm$^2$/s/Hz/sr. 

At high galactic latitudes and in regions of low $N_H$, the DGL contributes 
less than 26.0 mag arcsec$^{-2}$ at UV--optical wavelengths, or $\la 6\times
10^{-20}$erg/cm$^2$/s/Hz/sr (Henry 1991, O'Connell et al. 1992). 

The EBL 
surface brightness is estimated to be I$\ga$24--25 mag/arcsec$^2$, since the
galaxy counts are now known to permanently converge with a sub-critical 
magnitude slope $\alpha\la$0.4 for B$\ga$25 mag (Metcalfe et al. 1995) and for
I$\ga$24 mag (Williams et al. 1996). The best estimate of the extragalactic 
background is about 1.7$\times 10^{-20}$erg/cm$^2$/s/Hz/sr or 
$\sim$27.3 mag arcsec$^{-2}$ at $0.55\mu$m, and 6.7$\times 
10^{-20}$erg/cm$^2$/s/Hz/sr or $\sim$25.1 mag arcsec$^{-2}$ at $1\mu$m 
(Dwek et al. 1998; Hauser et al. 1998)

It is assumed here that the high-redshift intergalactic medium is not obscured
by dust and neutral clouds at redshifts somewhat lower than $z_{ion}$. 
Comparisons of radio and optical samples of high-redshift quasars (Shaver et
al. 1996, 1999) and the discovery of Ly$\alpha$ emitters at z$\simeq$5--6 (Dey
et al. 1998, Weymann et al. 1998) support this assumption. In any case, while
foreground dust could diminish the observed amplitude of any reionization
signal, it would not significantly affect its global spectral signature (c.f.,
Seaton 1979). 

Taking the above foreground emissions into account, the horizontal lines in
Fig. 5 show the sensitivity expected from a one-day integration with
HST/STIS (Space Telescope Imaging Spectrograph), HST/WFC3 (Wide Field 
Camera 3), NGST (the Next Generation Space Telescope), and a long integration 
on a similar spacecraft at 3 AU from the sun where the zodiacal
emission would be reduced by two orders of magnitude. The reionization
signal is within reach, at least for the deep space mission.

We note that in ground-based direct CCD-imaging local sky-removal to a
precision of $\sim 10^{-3}$ is obtained routinely, and that it has been
obtained to a precision better than $\sim 10^{-4}$ when using very careful
calibrations of the kind we propose below (e.g. Tyson 1988). There is no
fundamental reason why a CCD spectrograph outside the Earth's atmosphere
should not be able to achieve these limits in spectroscopic mode, although this
has yet to be demonstrated.

The sensitivity requirement may be eased by taking advantage of the
considerable spectral structure present in the curves shown in Fig. 5. Rather
than looking only for the Ly$\alpha$ and H$\alpha$ steps, it may be possible
to search for the overall pattern in these curves by cross-correlating
the observed spectrum with a
theoretical template (e.g. Baltz et al. 1998), as is often done in determining
redshifts of elliptical galaxies. This could ease the sensitivity 
requirement by
a considerable factor (which depends on the actual structure in the
reionization signal). In this case, however, the observable would only be a
cross-correlation peak rather than the actual spectral features, and such a
technique would require excellent flat-fielding and calibration over a wide
wavelength range.

\subsection{Contamination by galactic and extragalactic foregrounds} 

Any spectral features or variations in the foreground contaminants could
conceivably mask the reionization signal, but it is likely that such
contamination can be adequately dealt with. Deep ground-based imaging of each
field observed would identify any large-scale low-surface brightness objects
or structures to $\la$~29--30 mag arcsec$^{-2}$ (e.g. Tyson 1988) that can be
excluded afterwards. The observed sky background spectrum can then be corrected
for the diffuse galactic and extragalactic foreground emissions as follows:

(a) The zodiacal ($\simeq$solar) spectrum is the main contributor to the
variations in the spectral baseline, and is mostly featureless over the
wavelength range of interest, except for perhaps the solar Ca feature at
0.89$\mu$m. The solar spectrum is known to very high accuracy. The zodiacal
dust particles with sizes $<$1$\mu$m do not significantly change the high
frequency part of the zodiacal spectrum around 1$\mu$m, but they slightly
modify the global solar spectral gradient due to non-grey scattering. Such
changes are not expected to be sudden with wavelength. A slightly reddened
solar spectrum can be subtracted from the final observed sky spectrum after
appropriate scaling to fit the HST broad-band zodiacal sky-measurements. 

(b) Possible high spatial frequency structures in the galactic DGL at very low
surface brightness levels may be of some concern, but only at a level of $>$27
mag/arcsec$^2$, as has been seen in deep ground-based imaging at low galactic
latitudes (Szomoru \& Guhathakurta 1998; Guhathakurta \& Tyson 1989). At high
galactic latitudes, if the observed fields can be selected to have $E_{B-V} \la
0.01$ and/or low $N_{H}$, the DGL and variations therein should be very small,
as shown by deep ground-based images ({\it e.g.}, Tyson 1988) and deep HST
images such as the Hubble Deep Field (Williams et al. 1996). Fields near
potentially contaminating stars or galactic cirrus would have to be avoided.
Since the DGL likely bears the spectral signatures of early-type stars in our
Galaxy, we do not expect a large spectral feature around 1$\mu$m from the
combined stellar population that causes the reflected DGL. In any case, such a
signal can be subtracted from the residual sky spectrum at the B$\simeq$27 mag
arcsec$^{-2}$ level by subtracting a scaled OB star template, as for the
zodiacal spectrum. 

(c) Finally, the integrated background from faint foreground galaxies (EBL)
should be featureless, given their large redshift range (cf. Driver et al. 
1998). Therefore no global template for the EBL from faint galaxies at z~$\la
5$ should have to be subtracted from the collapsed spectrum to find a Ly-edge.

\subsection{Observational prospects}

\subsubsection{Observing the Lyman edge with space-based instruments } 

A near-infrared instrument in space would be essential for this experiment,
because of the low sky background (in space the zodiacal sky in the H-band is
6--7 mag darker than from the ground (Thompson et al. 1999; Windhorst \& 
Waddington 1999) and complete absence of terrestrial OH lines. We note that a
recent deep long-slit search with Keck provided quite useful upper limits to
the ionizing background $J_{\nu 0}$ at levels $2 \times 10^{-21}$
erg/cm$^2$/s/Hz/sr, but at lower redshifts (2.7$\la$z$\la$4.1; Bunker et al.
1998). In our case we are  concerned with redshifts z$\ga$5, and so we must
do the experiment from space to avoid the strong OH-forest. 

Ideally, one would do this experiment on a faster telescope with a larger
aperture and lower sky than HST, such as the NGST, to be launched sometime
after 2007. Improved infrared cameras with spectroscopic multiplexing 
capability, such as those under consideration for WFC3 and being designed 
for NGST, are also essential. They will have larger IR detectors with much 
better noise characteristics than HST/NICMOS, making possible a substantial 
gain in sensitivity for this experiment. The NGST will operate over the 
range $\lambda = $0.5--10$\mu$m, and will have a large spectroscopic 
multiplexing capability provided by an integral field or multi-object 
spectrograph with some 40,000 spectral pixels, One would search for the
isotropic step-like spectral features in the collapsed 1-dimensional spectrum
of the sky. 

Alternatively, an infrared space telescope dedicated to studying the CIBR
should be able to detect the Lyman edge signals. Two versions, EGBIRT and
DESIRE, have already been proposed (Mather \& Beichman 1996). These would
observe at 3 AU from the sun, reducing the zodiacal foreground by two orders
of magnitude.

The sensitivities attainable with such space-based facilities are
shown in Fig. 5. As indicated above, the level of the reionization signal can
be reached, but it may require the ``deep-space'' mission with its much lower
background and long dedicated integrations. In the meantime, an
exploratory experiment can be done with HST, as described below.

\subsubsection{An experiment with the HST} 

In view of the large uncertainties, both in the predicted amplitude of the
reionization signal and in the attainable sensitivities, it is worth
considering preliminary experiments using the currently existing Space
Telescope Imaging Spectrograph (STIS). Such experiments would also be
exploring new observational domains, and unexpected discoveries may result. For
example, in addition to the hydrogen recombination signal shown in Fig, 5,
there may also be a broader and possibly stronger spectral feature due to the
continuum emission from the ionizing sources near $z_{ion}$ as shown in fig.
11 of Gnedin \& Ostriker (1997). Such an experiment would in any case provide
important experience in making observations of this kind. 

The STIS has a wide variety of long slits, permitting a spectral image of the
sky to be obtained. To obtain sufficient surface brightness sensitivity, a long
observation with the widest available STIS long slit of 52''$\times$2.0''
should be made. One would use the longest-wavelength low-resolution STIS
grating G750L to make a deep spectrum of the sky. The observation would be
limited by the STIS CCD to $\la$~1.03$\mu$m, or z~$\la$~7.5 for Ly-$\alpha$. 

This HST observation could be done in parallel mode, at no extra cost in
primary HST observing time. As the primary exposures would most likely be taken
with WFPC2 and therefore be dithered, the STIS exposures would also be 
dithered along the slit, resulting in better spectral flatfields. Several
independent fields could be observed in order to identify and remove
``spurious'' features from foreground objects. One or two fields should be
taken at lower ecliptic latitudes -- or higher zodiacal foreground -- so that
the zodiacal spectrum could be removed better by comparison with the higher
latitude observations.

A project such as this relies on the long temporal stability and excellent
flat-field properties of STIS, and would push the STIS calibration to its 
limits. Contemporaneous calibrations would be essential to reduce the 
systematic errors in the spectra to an absolute minimum. The entire bright
side of each orbit should be used to make the maximum number of calibrations
between the parallels taken during the dark side of each orbit, without 
interfering with the primary observations: bias frames, dark frames, earth 
flats (done between the two darks in the middle of the bright side of the 
orbit, so that HST would be looking down for earth flats), and internal 
tungsten flats. 

The experiment would be read-noise limited, since one cannot expose for long
periods and still reject cosmic rays. Sensitivity could be gained by on-chip
binning, as the observations are not dark-current limited (Baum et al. 1998).
Fringe removal could be done with contemporaneous spectral flats (Baum et al.
1998). Following cosmic-ray removal ({\it c.f.}, Windhorst et al. 1994a), the
best possible bias, darks, and flats would have to be produced. The end goal
would be a spectral image that is globally flat-fielded to within
$\la~10^{-3}\times$ sky.

The following sensitivities can be reached in principle with the STIS/CCD/G750L
combination, according to the STIS Manual (Baum et al. 1998) and the STIS
exposure time calculator. With 36 orbits (total time 94 ksec) a 1$\sigma$
surface brightness sensitivity of $1.4\times 10^{-20}$ erg/cm$^2$/s/Hz/sr can
be achieved per pixel, or a 3$\sigma$ sensitivity of $1.3\times
10^{-21}$erg/cm$^2$/s/Hz/sr if the sky is flat to well within $\la 10^{-3}$, so
that one could collapse all 1000 spatial pixels along the slit into a
one-dimensional STIS sky spectrum. The approximate 3-$\sigma$ limit expected
from a 36 orbit HST/STIS integration with the grating G750L and the 2''
long-slit is shown in Fig. 5. The uncertainty in this STIS simulation is about
0.3 dex. 

The question then is whether any significant spectral structure survives in
the collapsed STIS spectrum after subtraction of the best-fit scaled 
templates of the zodiacal and DGL spectra, and after checking for possible 
coincidences with any of the weak known Ca features in the solar spectrum (or
perhaps in the DGL) around 0.89$\mu$m. If structure is not obvious, large 
wavelength swaths could be summed and moved in a sliding manner to find the 
most significant features. Alternatively the spectrum could be cross-correlated
with a theoretical template such as the curves in Fig. 5, or autocorrelated to
find any possible steps buried in the data. Such an experiment, pushing STIS
to its limits and exploring a new observational domain, will be an important
first step both technically and scientifically. 

 \section{Conclusions} 

The reionization of the Universe may be detectable globally in the form of a
step in the broadband spectrum of the sky. This would be a unique measurement
of a basic ``phase transition'' of the Universe. As such it would differ
fundamentally from previously proposed searches based on fluctuations due to
individual \ion{H}{i} concentrations or absorption against continuum sources
which may exist at still higher redshifts. The step is expected at $\sim$
70--240 MHz due to redshifted \ion{H}{i} 21 cm line emission and at $\lambda
\sim 0.7-2.6 \mu$m due to hydrogen recombination emission. The expected
amplitude is well above fundamental detection limits. The difficulties lie in 
contamination from galactic and extragalactic foregrounds, calibration, and
terrestrial interference. 

It appears that complexities due to the frequency dependence of the galactic
and extragalactic foregrounds may not be insurmountable obstacles to
identifying and measuring the reionization step. If that is the case, the
experiment is possible in principle, and the remaining hurdles are purely
technical. The challenge may be justified not only by the potential importance
of this measurement, but also by the possibility of opening up new domains in
observational parameter space.

\begin{acknowledgements}

We are grateful to Nick Gnedin and Ted Baltz for kindly providing their model
predictions, and we would like to thank Per Aannestad, Jaap Bregman, Frank
Briggs, John Mather, Don Melrose, Chris Salter, Tony Tyson, and Bruce Woodgate
for helpful discussions and comments.

\end{acknowledgements}

\end{document}